\documentclass[aps,pre,reprint]{revtex4-1} 
\usepackage{graphicx}
\usepackage{slashbox}
\usepackage{array}
\usepackage{color,epsfig}
\usepackage{fancyhdr,fancybox}
\usepackage{varioref,float,amssymb,amsmath}
\usepackage{float,amsfonts,amsthm}
\usepackage{changebar,longtable}
\usepackage[latin1]{inputenc}
\usepackage{fancyheadings}
\usepackage{makeidx}
\usepackage{psfrag}
\usepackage[below]{placeins}
\usepackage{flafter}
\usepackage[small]{caption2}

\newcommand{\bra}[1]{\left\langle{#1}\right\vert}
\newcommand{\ket}[1]{\left\vert{#1}\right\rangle}
\begin{document}

\title{Spatial Search Algorithms on Hanoi Networks}

\author{Franklin de Lima Marquezino}
\affiliation{Universidade Federal do Rio de Janeiro, RJ, 21941-972, Brazil}
\author{Renato Portugal}
\affiliation{Laborat\'{o}rio Nacional de Computa\c{c}\~{a}o Cient\'{\i}fica,
Petr\'opolis, RJ, 25651-075, Brazil}
\email{portugal@lncc.br}
\author{Stefan Boettcher}
\affiliation{Emory University, Atlanta, GA, 30322-2430, USA}

\date{\today}

\begin{abstract}
We use the \textit{abstract search algorithm} and its extension due
to Tulsi to analyze a spatial quantum search algorithm that finds a
marked vertex in Hanoi networks of degree 4 faster than classical
algorithms. We also analyze the effect of using non-Groverian coins
that take advantage of the small world structure of the Hanoi
networks. We obtain the scaling of the total cost of the algorithm
as a function of the number of vertices. We show that Tulsi's
technique plays an important role to speed up the searching
algorithm. We can improve the algorithm's efficiency by choosing a
non-Groverian coin if we do not implement Tulsi's method. Our
conclusion is based on numerical implementations.
\end{abstract}

\maketitle

\section{Introduction}

Grover's algorithm~\cite{Gro97a} allows one to find a marked item in
an unsorted database quadratically faster compared with the best
classical algorithm. It is the paradigm for many quantum algorithms
that use exhaustive search. The main technique used in Grover's
algorithm, called amplitude amplification, can be applied in many
computational problems providing gain in time complexity.

A related problem is to find a marked location in a spatial,
physical region. Benioff~\cite{Ben02} asked how many steps are
necessary for a quantum robot to find a marked vertex in a
two-dimensional grid with $N$ vertices. In his model, the robot can
move from one vertex to an adjacent one spending one time unit.
Benioff showed that a direct application of Grover's algorithm
does not provide improvements in the time complexity compared to a
classical robot, which is $O(N)$. Using a different technique,
called \textit{abstract search algorithms}, Ambainis
et.~al.~\cite{AKR05} showed that it is possible to find the marked
vertex with $O\left(\sqrt{N}\log N\right)$ steps. Tulsi~\cite{Tul08}
was able to improve this algorithm obtaining the time complexity
$O\left(\sqrt{N\log N}\right)$.

The time needed to find a marked vertex depends on the spatial
layout. The abstract search algorithm is a technique that can be
applied to any regular graph. It is based on a modification of the
standard discrete quantum walk. The coin is the Grover operator for
all vertices except for the marked one which is $-I$. The choice of
the initial condition is also essential. It must be the uniform
superposition of all states of the computational basis of the
coin-position space. This technique was applied with success on
higher dimensional grids~\cite{AKR05}, honeycomb
networks~\cite{ADMP10}, regular lattices~\cite{HT10} and triangular
networks~\cite{ADFP12}. Spatial search in Hanoi network of degree 3
(HN3) was analyzed in Ref.~\cite{MPB11}. Recently, the abstract
search algorithm was applied for spatial search on Sierpinski
gasket~\cite{PR12}.

In this work, we analyze spatial search algorithms
on the Hanoi network of degree 4 (HN4) extending
the analysis performed for HN3~\cite{MPB11}. HN4 is a
special case of \textit{small world networks}, which are being used
in many contexts including quantum computing~\cite{GGS05,MPB07}. We
also analyze the use of a modified coin operator instead of the Grover
coin, which is used in the standard form of the abstract search
algorithm,  to take advantage of the small world structure. Our
results are based on numerical simulations, but the hierarchical
structure of HN4 indicates that analytical results can also be
obtained. Hanoi networks have a hierarchical structure of fractal
type that helps to gain insights of spatial search algorithms in
graphs that are not translational invariant. Recently, there has
some effort in this direction~\cite{ABM10, PR12}

The structure of the paper is as follows. Sec.~\ref{sec:HS}
introduces the degree-4 Hanoi network. Sec.~\ref{sec:QW} describes
the standard coined discrete quantum walk on HN4. Sec.~\ref{sec:ASA}
reviews the basics of the abstract search algorithm and Tulsi's
method. Sec.~\ref{sec:MM} describes the modification on the coin
operator we propose to enhance the time complexity of quantum search
algorithms on HN4. Sec.~\ref{sec:Results} describes the main results
based on numerical simulations. Finally, we present our final remarks
in Sec.~\ref{sec:Conclusions}.

\section{Hierarchical Structures}
\label{sec:HS}

\begin{figure}[h]
    \centering
    \includegraphics[height=5.0cm]{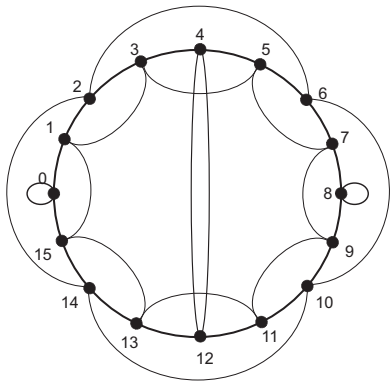}
    \caption{HN4 with 16 vertices.}
    \label{fig:NH4}
\end{figure}

The Hanoi network has a cycle with $N=2^n$ vertices as a backbone
structure, that is, each vertex is adjacent to 2 neighboring
vertices in this structure and extra, long-range edges are
introduced with the goal of obtaining a small-world hierarchy. The
labels of the vertices $0< k \le 2^n-1$ can be factorized as
\begin{equation}\label{eq:k}
    k =2^{k_1} (2\,{k_2} + 1),
\end{equation}
where ${k_1}$ denotes the level in the hierarchy and ${k_2}$ labels
consecutive vertices within each hierarchy. In any level, one links
the vertices with consecutive values of ${k_2}$ keeping the degree
constraint. When ${k_1}=0$, the values of $k$ are the odd integers.
For HN4, we link 1 to 3, 3 to 5, 5 to 7 and so on. The vertex with
label 0, not being covered by Eq.~(\ref{eq:k}), has a loop as well as
vertex of label $2^{n-1}$. Fig.~\ref{fig:NH4} shows all edges for
HN4 when the number of vertices is $16$.  Using this figure, one can
easily build HN4 recursively, each time doubling the number of
vertices. Our analysis will be performed for a generic value of $n$
to allow us to determine the computational cost as function of $N$.

HN4 has a small-world structure because the diameter of the network
only increases with $\sim\sqrt{N}$~\cite{BGA08},  less fast than the
number of vertices $N$. Yet, HN4 is a regular graph of fixed degree
$d=4$ at each vertex.

\section{Quantum Walks on Hierarchical Structures}\label{sec:QW}

A coined quantum walk in HN4 with $N=2^n$ vertices has a Hilbert
space ${\cal H}={\cal H}_C\otimes {\cal H}_P$, where ${\cal H}_C$ is
the $4$-dimensional coin subspace and ${\cal H}_P$ the
$N$-dimensional position subspace. A basis for ${\cal H}_C$ is the
set $\{\ket{a}\}$ for  $0\leq a\leq 3$ and ${\cal H}_P$ is spanned
by the set $\{\ket{k}\}$ with, $0 \leq k\leq N-1$. We use the
decomposition $k=(k_1,k_2)$, given by Eq.~(\ref{eq:k}), when
convenient. A generic state of the discrete quantum walker in HN4 is
\begin{equation}
\ket{\Psi(t)}=\sum_{a=0}^{3}\sum_{k=0}^{N-1}\psi_{a,k}(t)\ket{a}\ket{k}.
\label{eq:estgeral1d}
\end{equation}

The evolution operator for the standard quantum walk~\cite{Kem03} is
\begin{equation}\label{evol}
U=S\circ(C\otimes I),
\end{equation}
where $I$ is the identity in ${\cal H}_P$ and $S$ is the shift
operator defined by the following equations,
\begin{eqnarray*}
  S\ket{0}\ket{k} &=&
  \begin{cases}
    \ket{1}\ket{k_1,k_2+1}, & \mbox{ if } k_2 \mbox{ is even}\\
    \ket{1}\ket{k_1,k_2-1}, & \mbox{ if } k_2 \mbox{ is odd}\\
  \end{cases},\\
  S\ket{1}\ket{k} &=&
  \begin{cases}
    \ket{0}\ket{k_1,k_2+1}, & \mbox{ if } k_2 \mbox{ is even}\\
    \ket{0}\ket{k_1,k_2-1}, & \mbox{ if } k_2 \mbox{ is odd}\\
  \end{cases},
\end{eqnarray*}
and
\begin{eqnarray*}
  S\ket{0}\ket{2^{n-1}} &=& \ket{1}\ket{2^{n-1}}, \\
  S\ket{1}\ket{2^{n-1}} &=& \ket{0}\ket{2^{n-1}}, \\
  S\ket{0}\ket{0}       &=& \ket{1}\ket{0}, \\
  S\ket{1}\ket{0}       &=& \ket{0}\ket{0}, \\
  S\ket{2}\ket{k}       &=& \ket{3}\ket{k+1}, \\
  S\ket{3}\ket{k}       &=& \ket{2}\ket{k-1}.
\end{eqnarray*}
The arithmetical operations on the second ket is performed modulo
$N$. The shift operator obeys $S^2=I$.  $C$ is a unitary coin
operation in ${\cal H}_C$. In the standard walk, $C$ is the Grover
coin, denoted by $G$,
\begin{equation}
    G=\frac{1}{2}
    \begin{pmatrix}
    -1&1 &1 & 1\\
    1 &-1&1 & 1\\
    1 &1 &-1& 1\\
    1 &1 &1 &-1
    \end{pmatrix}
\end{equation}
which is the most diffusive coin~\cite{NV00}.

The dynamics of the standard quantum walk is given by
\begin{equation}\label{eq:U_t}
   \ket{\Psi(t)}= U^t \ket{\Psi_0},
\end{equation}
where $\ket{\Psi_0}$ is the initial condition. After $t$ steps of unitary evolution, we perform a position measurement which yields a probability distribution given by
\begin{equation}\label{eq:PROB  }
    p_k = \sum_{a=0}^2 \left| \bra{a,k}U^t \ket{\Psi_0} \right|^2.
\end{equation}

\section{Abstract Search Algorithms}\label{sec:ASA}

The \textit{abstract search algorithm}~\cite{AKR05} is based on a
modified evolution operator $U^\prime=S \cdot C^\prime$, obtained
from the standard quantum walk operator $U$ by replacing the coin
operation $C$ with a new unitary operation $C^\prime$ which is not
restricted to ${\cal H}_C$ and acts differently on the searched
vertex. The modified coin operator is
\begin{equation}\label{eq:Cprime}
    C'= -I\otimes \ket{k_0}\bra{k_0} + C\otimes (I- \ket{k_0}\bra{k_0}),
\end{equation}
where $k_0$ is the marked vertex in a regular graph and $C$ is the
Grover coin $G$, the dimension of which depends on the degree of the
graph. Ambainis et.~al.~\cite{AKR05} have shown that the time
complexity of the spatial search algorithm can be obtained from the
spectral decomposition of the evolution operator $U$ of the
unmodified quantum walk, which is usually simpler than that of $U'$.

The initial condition $\ket{\psi_0}$ is the uniform superposition of
all states of the computational basis of the whole Hilbert space.
This can be written as the tensor product of the uniform
superposition of the computational basis of the coin space with the
uniform superposition of the position space. Usually, this initial
condition can be obtained in time $O(\sqrt N)$, where $N$ is the
number of vertices.

The evolution operator is applied recursively starting with the
initial condition $\ket{\psi_0}$. If $t_f$ is the running time of
the algorithm, the state of the system just before measurement is
$U'^{t_f}\ket{\psi_0}$. If one analyzes the probability of obtaining
the marked vertex $k_0$ as function of time since the beginning of
the algorithm, one gets an oscillatory function with the first
maximum close to $t_f$.

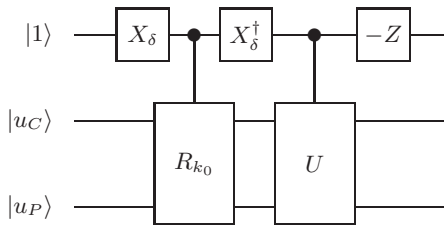
\begin{figure}[h]
  \centering
  \setlength{\unitlength}{0.65pt}
  \begin{picture}(160,160)(80,0)
    \put(43,25){\makebox(0,0)[r]{$\vert{u_{P}}\rangle$}}
    \put(250,125){\line(1,0){20}}
    \put(80,110){\framebox(30,30){$X_{\delta}$}}
    \put(140,110){\framebox(30,30){$X_{\delta}^{\dagger}$}}
    \put(220,110){\framebox(30,30){$-Z$}}
    \put(218,25){\line(1,0){52}}
    \put(218,75){\line(1,0){52}}
    \put(102,15){\framebox(46,70){\small{$R_{k_0}$}}}
    \put(148,25){\line(1,0){24}} \put(148,75){\line(1,0){24}}
    \put(195,125){\circle*{8}} \put(195,121){\line(0,-1){36}}
    \put(172,15){\framebox(46,70){$U$}}
    \put(43,125){\makebox(0,0)[r]{$\vert{1}\rangle$}}
    \put(43,75){\makebox(0,0)[r]{$\vert{u_{C}}\rangle$}}
    \put(125,125){\circle*{8}}
    \put(170,125){\line(1,0){50}}
    \put(110,125){\line(1,0){30}}
    \put(55,125){\line(1,0){25}}
    \put(125,121){\line(0,-1){36}}
    \put(55,25){\line(1,0){47}}
    \put(55,75){\line(1,0){47}}
  \end{picture}
  \caption{Tulsi's circuit diagram for the one-step evolution operator of the quantum walk search algorithm. $\ket{u_P}$ is the uniform superposition of the computational basis of the position space.}\label{fig:circ}
\end{figure}

After a little algebra, the evolution operator $U^\prime=S \cdot
C^\prime$ can be converted into the form  $U'=U\cdot R_{k_0}$, where
$R_{k_0}=I-2\ket{u_C,k_0}\bra{u_C,k_0}$, $U$ is given by
Eq.~(\ref{evol}), and $\ket{u_C}$ is the uniform superposition of
the computational basis of the coin space. Using the expression
$U'=U\cdot R_{k_0}$ as a starting point, Tulsi proposed a new
version of the search algorithm, which requires an extra register
(an ancilla qubit) used as a control for the operators $R_{k_0}$ and
$U$. The operators acting on the ancilla register are described in
Figure~\ref{fig:circ}, where $-Z$ is the negative of Pauli's $Z$
operator and
\begin{equation}\label{eq:Xdelta}
  X_{\delta}= \left( \begin{array}{cc} \cos\delta & \sin\delta \\ -\sin\delta & \cos\delta\end{array} \right).
\end{equation}
The value of $\delta$ is the one that optimize the cost of the
algorithm. For two-dimensional lattices, Tulsi~\cite{Tul08} showed
that $\cos \delta \propto 1/\sqrt{\log N}.$

The Tulsi's evolution operator is
\begin{equation}\label{eq:new_U}
    U''=(-Z\otimes I)\cdot C(U)\cdot (X_{\delta}^{\dagger}\otimes I) \cdot C(R_{k_0}) \cdot (X_{\delta}\otimes I),
\end{equation}
where $C(U)$ and $C(R_{k_0})$ are the controlled operations shown in
Figure~\ref{fig:circ} and $I$ is the identity operator in ${\cal
H}$. We want to determine how many times $U''$ must be iterated,
taking $\ket{1}\ket{u_C}\ket{u_P}$ as the initial condition, in
order to maximize the overlap with the search element.

\section{Modified Method}\label{sec:MM}

The coin in a quantum walk is used to determine the direction of the
movement. The Grover coin is an isotropic operator regarding
all outgoing edges from a vertex. It is useful in networks that have
no special directions, such as two-dimensional grids and hypercubes.
The Hanoi network, on the other hand, has a special direction that
creates the small world structure. Any edge that takes the walker
outside the circular backbone provides an interesting opportunity in
terms of searching. The strategy is to have a parameter that can
control the probability flux among the edges, reinforcing or
decreasing the flux outwards or inwards the circular backbone.

Instead of using the Grover coin of the \textit{abstract search
  algorithms}, we analyze the use of a modified coin given by
\begin{equation}\label{eq:new_C}
  \begin{split}
    C = \frac{2\epsilon}{d}\left(\ket{0}\bra{0} +
            \ket{1}\bra{1} +
            \ket{0}\bra{1} + \ket{1}\bra{0} \right) +\\
             2\sqrt{\frac{\epsilon(d-2\epsilon)}{d^2(d-2)}}\sum_{j=2}^{d-1}\left(\ket{j}\bra{0} + \ket{0}\bra{j}\right) +\\
             2\sqrt{\frac{\epsilon(d-2\epsilon)}{d^2(d-2)}}\sum_{j=2}^{d-1}\left(\ket{j}\bra{1} + \ket{1}\bra{j}\right) +\\
        \frac{2(d-2\epsilon)}{d(d-2)}\sum_{j,j'=2}^{d-1}\left(\ket{j}\bra{j'} \right) - I,
  \end{split}
\end{equation}
where $d=4$ is the degree at each vertex. When $\epsilon=1$, the
Grover coin is recovered. When $0<\epsilon<1$, the probability flux
along small-world edges (labels 0 and 1) which escapes from the
circular backbone is weakened.  When $1<\epsilon<4$, the probability
flux off the backbone is reinforced, allowing the walker to use the
small world structure with higher efficiency. Hence, this new coin
controls the bias to escape off the circular backbone
  of HN4 through the parameters $\epsilon$.

The \textit{abstract search algorithms} use a uniform distribution
as initial condition. We change this recipe. The initial condition
is
\begin{equation}\label{eq:initial_condition}
\begin{split}
  \ket{\psi(0)}=\sqrt{\frac{\epsilon}{d}}\big(\ket{0}\ket{s} + \ket{1}\ket{s}\big) +\\
                \sqrt{\frac{d-2\epsilon}{d(d-2)}}\sum_{j=2}^{d-1}\ket{j}\ket{s},
\end{split}
\end{equation}
where $\ket{s}$ is the uniform superposition on the position space.
When $\epsilon=1$ the initial condition is the uniform superposition
of coin-position space.

We want to check whether it is possible to improve the abstract
search algorithm by modifying the coin operator for HN4 in such way
we can tune parameter $\epsilon$ for obtaining the best rate of
probability flux between the circle backbone and small-world edges.
In a previous paper~\cite{MPB11}, we have concluded that, for HN3
without using Tulsi's method, it is better to choose
$\epsilon=1.75$. After analyzing further this issue, we have to
reconsider this conclusion, mainly when one uses Tulsi's method,
which seems to favor the Grover coin even in nonhomogeneous graphs.

In this work, we consider three different methods: 1) the abstract
search algorithm, 2) the Tulsi's method, and 3) the modified method.
The analysis of the evolution of the quantum search algorithm using
the new coin and initial condition is far more complex than the
standard one. Our conclusions here are based in numerical
simulations.

\section{Main Results}\label{sec:Results}

\begin{figure}[!h]
  \centering
  \includegraphics[width=0.9\columnwidth]{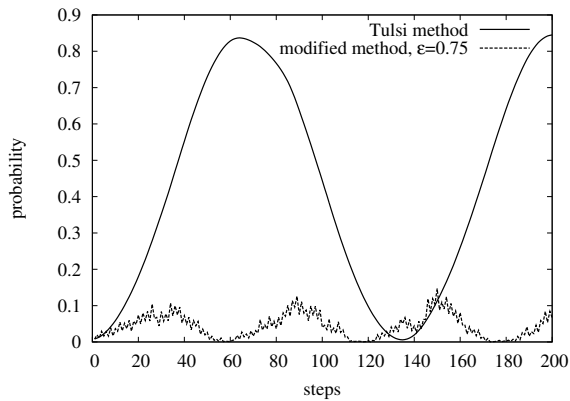}
  \caption{Probability of finding walker at vertex $k=3$ as a function
    of time using the modified method and the abstract search algorithm
    with Tulsi's method.}
  \label{fig:prob-vs-t}
\end{figure}

Fig.~\ref{fig:prob-vs-t} shows the oscillatory behavior of the
probability of finding the walker at the marked vertex using the
modified method with $\epsilon=0.75$ (lower curve) and the
\textit{abstract search algorithm} using Tulsi's method (higher
curve). Initially, the probability is close to zero, because the
initial condition is a state that is close to the uniform
superposition of all vertices. The running time of the algorithm is
the value of $t$ for which the probability is close to its first
maximum. Note that, without using Tulsi's method, the maximum value
of the probability is smaller than that of the abstract search
algorithm with Tulsi's method. In either case, the maximum value of
the probability is not close to 1, as one would expect in order to
have high probability to find the marked vertex. This means that the
algorithm must be rerun many times to amplify the success
probability. For the lower curve, the number of repetitions is
large, in fact, it scales with $N$, which has a strong impact on the
total cost of the algorithm. We call \textit{success probability}
the value of the probability for the first peak of
Fig.~\ref{fig:prob-vs-t}. The marked vertex used in
all simulations is $k=3$, but the conclusions will not
depend on the hierarchy of the target vertex.

\begin{figure}[!h]
  \centering
  \includegraphics[width=0.9\columnwidth]{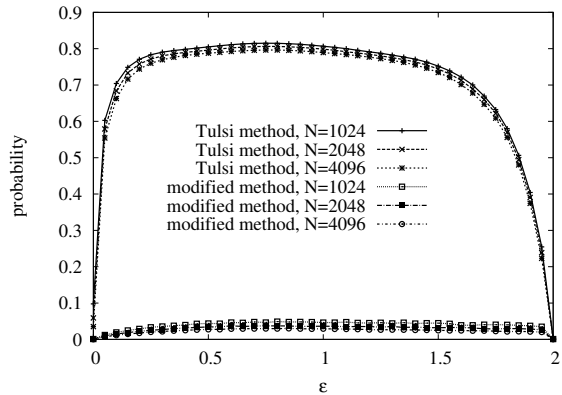}
  \caption{Success probability of the search algorithm as function
    of $\epsilon$ for three values of $N$ using both the modified coin and
    Tulsi's method on the top of the abstract search algorithm.}
  \label{fig:prob-vs-eps-HN4}
\end{figure}

Now let us try to answer the following question about parameter
$\epsilon$: What is the best value of $\epsilon$ for the spatial
search algorithm? Fig.~\ref{fig:prob-vs-eps-HN4} shows the success
probability as function of $\epsilon$ for three values of $N$ both
for the modified method (lower curves) and the abstract search
algorithm using Tulsi's method (higher curves). The curves are very
flat around $\epsilon=1$, which correspond to the Grover coin. This
shows the the modified method does not play an important role for
improving the efficiency of the algorithm. These curves do not
provide enough clues for choosing $\epsilon$. The final answer can
be achieved by analyzing the effect of $\epsilon$ on the total cost
of the algorithm. We measure the cost as the number of times the
evolution operator is applied, or equivalently the number of oracle
queries considering the repetitions necessary for amplitude
amplification.

\begin{figure}[!h]
  \centering
  \includegraphics[width=0.9\columnwidth]{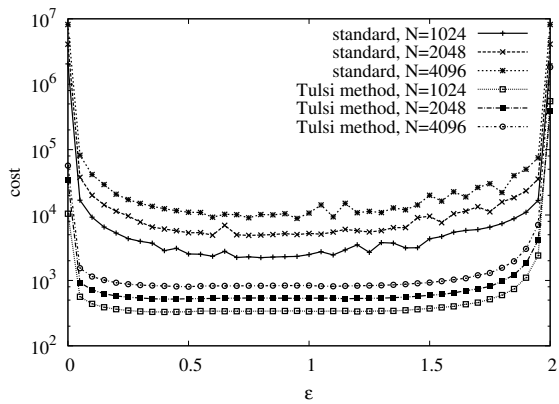}
  \caption{Computational cost of the search algorithm in terms of
    $\epsilon$ for two values of $N$ using both the modified coin and Tulsi's method on the top
    the abstract search algorithm.}
  \label{fig:cost-vs-N-NH4}
\end{figure}

Fig.~\ref{fig:cost-vs-N-NH4} shows the total cost of the search
algorithm as function of $\epsilon$ for two values of $N$ both for
the modified method (higher curves) and the abstract search
algorithm using Tulsi's method (lower curves). The curves are very
flat around $\epsilon=1$ as before, but we can conclude that the
best values are $\epsilon=0.75$ for the modified method and
$\epsilon=1$ (Grover coin) for the abstract search algorithm using
Tulsi's method. From now on, we will take these values of $\epsilon$
for the rest of this paper. The fact that the algorithm cannot be
improved by modifying the coin after Tulsi's method seems to show
that the algorithm has achieved its best performance using the Grover
coin. This conclusion for a graph that has non-homegeneous vertices
and non-isotropic edges was not the one expected by us at the
beginning and it is quite surprising.

\begin{figure}[!h]
  \centering
  \includegraphics[width=0.9\columnwidth]{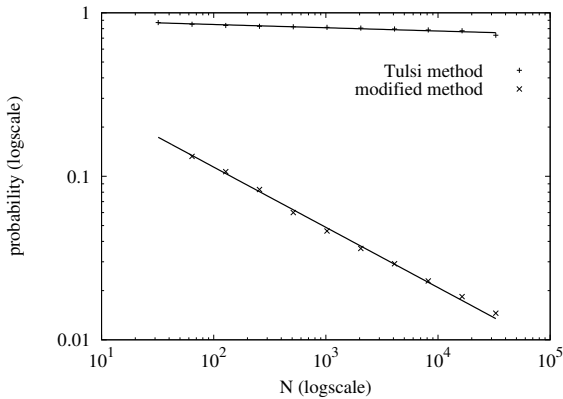}
  \caption{Success probability of the search algorithm as a function
    of $N$. The fitting shows that $P=0.62/N^{0.37}$ for the modified method.}
  \label{fig:prob-vs-N}
\end{figure}

Fig.~\ref{fig:prob-vs-N} shows the success probability after a
single run of the search algorithm as a function of the network size
$N$ in log-scale for the modified method (lower points) and the
abstract search algorithm using Tulsi's method (higher points). From
the inclination of the best fitting line, we conclude that the
success probability of the modified method decays approximately as
$0.62/N^{0.37}$. Using the technique of amplitude
amplification~\cite{AKR05}, the algorithm must be rerun around
$O(N^{0.185})$ times in order to ensure a final probability close to
$1$. This produces a high impact on the total cost of the algorithm.
Recall that for the two-dimensional grid, the number of repetitions
is $O(\sqrt{\log N})$~\cite{AKR05}. This result shows that Tulsi's
method plays an important role in terms of computational complexity.
The best value of $\delta$ in Eq.~(\ref{eq:Xdelta}) for HN4 seems to
be $\cos\delta=O(1/\log N)$. In Fig.~\ref{fig:prob-vs-N}, we see
that success probability after a single run of the abstract search
algorithm using Tulsi's method does not depend on $N$. In this case,
we can rerun the algorithm a fixed number of times to obtain an
overall probability very close to 1.

\begin{figure}[!ht]
  \centering
  \includegraphics[width=0.9\columnwidth]{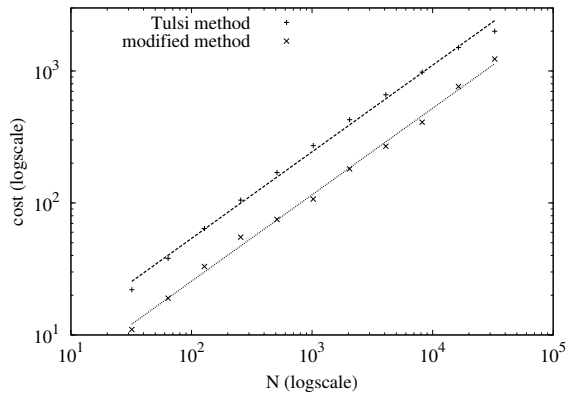}
  \caption{Computational cost of the search algorithm in terms of
    $N$ in log scale. The dotted curves are fitting curves.
    For the modified method, the best fitting is $1.25 N^{0.65}$.
    For the Tulsi method, the best fitting is $2.62 N^{0.65}$.}
  \label{fig:cost-vs-N}
\end{figure}

Fig.~\ref{fig:cost-vs-N} shows the computational cost of the search
algorithm as a function of the network size $N$ both for the
modified method (cross points) and the abstract search algorithm
using Tulsi's method (x points). We have not used the method of
amplitude amplification in this experiment. We have displayed the
best fitting lines for both cases, which scale as $O(N^{0.65})$. For
the Tulsi method, this is the total cost of the algorithm, because
the success probability is high (see Fig.~\ref{fig:prob-vs-N}). For
the modified method, the total cost is $O(N^{0.84})$, because we
must use the method of amplitude amplification, which puts an
overhead of $1/\sqrt P$, where $P=O(1/N^{0.37})$ (see
Fig.~\ref{fig:prob-vs-N}).

It is important to note that the data in Fig.~\ref{fig:cost-vs-N}
is not precise enough to detect the presence of $\log N$ terms
in the expression of the total cost. We are using results with
small $N$ to draw conclusions about the asymptotic behavior.
There are imprecisions in the simulations that come from
the discrete nature of the spatial layout. For instance,
the lower curve in Fig.~\ref{fig:prob-vs-t} has quick
oscillations that have some impact on the
impreciseness of the total cost. Usually,
the results using Tulsi's method are more stable.

The scaling of the total cost has a slight variation when
we change the position of the target. On the other hand,
the prefactor changes. In terms of graph structure,
HN4 is non-homogeneous, because the vertices can be divided in
hierarchical levels. This non-homogeneity does not play an
important role in term of the cost of finding a vertex. The
same kind of conclusions holds for HN3, analyzed in Ref.~\cite{MPB11}.

We have redone and extended the simulations for HN3, using
a new implementation. The scaling for the total cost in terms
of $N$ is (1) $O(N^{0.62})$ using Tulsi's method, and (2) $O(N^{0.74})$
using the modified method with $\epsilon=1.75$ applying amplitude
amplification. There are two important conclusions we draw from
the comparison between HN3 and HN4, one is about the graph degree
and the other about the value of $\epsilon$.

The degree of the
graph seems to play no important role in terms of efficiency.
Similar conclusions were draw in Ref.~\cite{ADFP12}, which
compared the efficiency of quantum search algorithms in
triangular, square, and hexagonal lattices that have
degrees 3, 4, and 6, respectively. The scaling of the cost
as a function of $N$ is the same for all of them.

The optimal value of $\epsilon$ for HN3
is larger than 1 using
the modified method, which means that the probability flux toward
the edges leaving the circle backbone is enhanced. For HN4,
the optimal value of $\epsilon$ is 0.75, smaller than 1.
We cannot say that long range connections (in term of
hierarchical level) helps in the quantum search. In fact,
for HN4 we have to decrease the probability flux in the
edges leaving the circle backbone. This can
be interpreted, when we take into account that HN4 is really
small-world and mean-field like in terms of the average distance
between any two vertices, which scales logarithmically
with system size, whereas the average distance scales
as $\sqrt{N}$ for HN3.
That means, in HN4 it is less significant to take long-range jumps,
because a random mix is already enough to get to most other sites;
whereas in HN3, if the walker does not take more long-range jumps than
nearest-neighbor jumps, it is difficult to go very far.

\section{Final Remarks}\label{sec:Conclusions}

We have analyzed spatial search algorithms on degree-4 Hanoi
networks with the goal of extending the abstract-search-algorithm
technique for nonhomogeneous graph structures with fractal nature.
We have proposed a
modification of the abstract search algorithm by choosing a coin
that takes advantage of the edge asymmetry of HN4. We have obtained
a faster algorithm by tuning numerically parameter $\epsilon$. The
cost of this algorithm is $O(N^{0.84})$ in terms of number of oracle
queries. The algorithm uses the standard method of amplitude
amplification on top of the modified abstract search algorithm with
$\epsilon=0.75$. This value of $\epsilon$ tells us that the probability
flux is higher on the circle backbone of HN4 than on the edges that
produces the small world structure.

We have also analyzed Tulsi's method on top of the abstract search
algorithm. In this case, the Grover coin ($\epsilon=1$) seems to be the best option and
the modified method does not improve the algorithm. The cost of the
algorithm is $O(N^{0.65})$. This is above the lower bound, which is
$O(N^{0.5})$, and above the cost of searching a marked vertex
on two-dimensional lattices, which is $O((N\log N)^{0.5})$. We have used
numerical methods to estimate the cost scale. This means that $\log N$ factors
may be lost, which could decrease the scale of $0.65$ in the total cost.

Our works on progress are now focused on obtaining analytical results
regarding search algorithms and general quantum walks for the Hanoi network HN4.

\end{document}